\documentclass{article}
\usepackage{graphics}
\usepackage{graphics}
\usepackage{amsfonts}
\usepackage{amssymb}
\usepackage{amsmath}
\newcommand{\tr}[1]{{\rm tr }\left(#1\right)}   

\newcommand{\set}[1]{\left\{#1\right\}} 

\newcommand{\flind}[1]{\mathfrak{L}\kern-8pt{-}(#1)}
\newcommand{\preflind}[1]{\mathfrak{L}_*\kern-12pt{-}\;\;(#1)}

\newcommand{\bo}[1]{\mathfrak{L}(#1)}

\newcommand{\1}{\mathbf{1}}
\newcommand{\vN}[1]{\mathcal{#1}}

\newcommand{\C}[1]{{\mathbb C}^{#1}}

\newcommand{\adj}[1]{{#1}^*}

\newcommand{\conj}[1]{{#1}^{\#}}
\newcommand{\cpos}[1]{\mathcal{B}^+(#1)}

\newtheorem{thm}{Theorem}

\newenvironment{pf}{\noindent{\it Proof.} }{\;\fbox{}\\}

\numberwithin{equation}{section}

\title{On completely positive non-Markovian evolution of a d-level system}
\author{Andrzej Kossakowski\\{\footnotesize\it Institute of Physics}\\{\footnotesize\it Nicholaus Copernicus University}\\{\footnotesize\it 87-100 Torun, Poland}\\{\footnotesize\it kossak@fizyka.umk.pl}\\ and \\Rolando Rebolledo \footnote{Partially supported by PBCT-ACT13 and Direcci\'on de Relaciones Internacionales-PUC}\\{\footnotesize\it Laboratorio de An\'alisis Estoc\'atico}\\{\footnotesize\it Facultad de
Matem\'aticas}\\{\footnotesize\it Pontificia Universidad Cat\'olica de Chile}\\{\footnotesize\it Casilla 306, Santiago 22,
Chile}\\{\footnotesize\it rrebolle@puc.cl}}
\date{}
\begin{document}
\maketitle
\begin{abstract}
A sufficient condition for non-Markovian master equation which ensures the complete positivity of the resulting time evolution is presented.\end{abstract}
\section{Introduction}
An open system is one coupled to an external environment \cite{Alicki:1987yo,Breuer:2002uq}. The interaction between the system and its environment leads to phenomena of decoherence and dissipation, and for this reason recently receive intense consideration in quantum information, where decoherence is viewed as a fundamental obstacle to the construction of quantum information processors\cite{Nielsen:2000ly} . In principle, the von Neumann equation for the total density matrix of the system and the reservoir provides complete predictions for all the observables. However, this equation is in practice impossible to solve since all degrees of freedom of the reservoir have to be taken into account. Main efforts have focused in deducing the time evolution of the reduced state density matrix. This is the aim of the well-known exact theory of subsystem dynamics due to Nakajima-Zwanzig (\cite{Nakajima:1958fx,Zwanzig:1960zh}) which relies in a generalized (non-Markovian) master equation approach.

The Nakajima-Zwanzig projection operator method makes possible to derive an exact equation for the reduced density from the von Neumann equation of the composed system. The resulting generalized master equation -an integrodifferential equation- is mostly of formal interest since such an exact equation can almost never be even written down explicitely in the closed form. In contrast, when one makes the Markovian approximation, i.e., when one neglects the reservoir memory effects, the resulting Markovian master equation \cite{Gorini:1976ta, Lindblad:1976ya} takes a simple form and the required \cite{Kraus:1983ta} complete positivity of the resulting time evolution is maintained. The main goal of the theory of open quantum systems is a non-Markovian description of the dynamics which at the same time include reservoir memory effects and retain complete positivity.

A variety of non-Markovian master equations have been proposed (cf. \cite{Breuer:2002uq,Shibata:1977mi,Imamoglu:1994pi,Royer:1996ff,Royer:2003lh,Barnett:2001fu,Breuer:2001qy,Breuer:2001kh,Breuer:1999rq,Breuer:2004fc,Breuer:2004ss,Breuer:2006gf,Maniscalco:2005ye,Maniscalco:cq,Wilkie:2000qo,Wilkie:2001tw,Breuer:qc,Breuer:2007eg,Budini:2004jl,Budini:il,Shabani:2005gb,Lee:2004mb,Pulo:fm}). However, the complete positivity of the resulting time evolution is still an important problem to be investigated.

In the present paper a sufficient condition for non-Markovian master equations is given, which ensures that the resulting time evolution is completely positive.
It is shown that this condition is rather difficult to verify in practice. The main reason for that is related to the normalization condition of the time evolution. This difficulty can be overcomed if one looks first for completely positive unnormalized solutions to non-Markovian master equations, while the normalization is imposed separately.
\section{Notations}
Let $\C{d}$ be a $d$-dimensional Hilbert space with the scalar product $\langle\cdot,\cdot\rangle$ and elements $e,x,y,z,\ldots$.

The $C^*$--algebra of linear operators on $\C{d}$ will be denoted by $M_{d}$. Elements of $M_{d}$ will be denoted by $a,b,c,\ldots$ and the unit of $M_{d}$ is $\1_{d}$. The $M_{d}$ is the Hilbert space under the scalar product $\langle a,b\rangle=\tr{\adj{a}b}$.

The $C^*$--algebra of linear maps from $M_{d}$ into $M_{d}$ will be denoted by $\bo{M_{d}}$, its elements are $A,B,C,\ldots$ and the identity map in $\bo{M_{d}}$ will be denoted by $id$. The conjugation (duality) $\conj{\cdot}$ in $\bo{M_{d}}$ is defined by the relation:
\begin{equation}
\langle \conj{A}a,b\rangle=\langle a,Ab\rangle,
\end{equation}
for all $a,b\in M_{d}$.

This operation endows the following property: the relations

\begin{equation}
A\1_{d}=\1_{d},\;\;L\1_{d}=0,
\end{equation}
and
\begin{equation}
\tr{\conj{A}a}=\tr{a},\;\;\tr{\conj{L}a}=0,
\end{equation}
are equivalent.

The cone of all completely positive maps on $M_{d}$ will be denoted by $\cpos{M_{d}}$.

Finally, if $A_{t}\in\bo{M_{d}}$, $t\geq 0$, then the Laplace transform of $A_{t}$ will be denoted by $\hat{A}_{p}$.
\section{Non-Markovian master equations}
The reduced dynamics can be studied equivalently in the Schr\"odinger or the Heinsenberg pictures. Suppose that $A_{t}:M_{d}\to M_{d}$ describes the reduced dynamics in the Heisenberg picture, then it should satisfy the following conditions:
$A_{t}\in\cpos{M_{d}}$, $A_{t}\1_{d}=\1_{d}$, for all $t\geq 0$, and $A_{0}=\lim_{t\to 0}A_{t}=id$. In the Schr\"odinger picture these relations are given in terms of $\conj{A}_{t}$, $t\geq 0$.

In the present section, the reduced dynamics is investigated under the assumption that $A_{t}$ is the solution of a non-Markovian master equation of the form:
\begin{equation}\label{nonmarkovian}
\frac{dA_{t}}{dt}=LA_{t}+\int_{0}^tdsL_{t-s}A_{s},
\end{equation}
with the initial condition $A_{0}=id$, where
\begin{equation}\label{3.2}
La=i[h,a]+Fa-\frac{1}{2}\set{F(\1_{d}),a},
\end{equation}
and $h=\adj{h}\in M_{d}$, $F\in\vN{B}^+(M_{d})$, that is, $L$ is the generator of a completely positive semigroup.

The normalization condition $A_{t}\1_{d}=\1_{d}$. implies the equality
\begin{equation}
L_{t}\1_{d}=0.
\end{equation}

A non-Markovian master equation of the form \eqref{nonmarkovian}can be easily derived from the Heisenberg equation for the composed system by the Nakajima-Zwanzig method under the assumption of factorization of the initial state of the composed system and the invariance of the initial reservoir state under the reservoir free evolution, c.f. \cite{Breuer:2002uq}. In this case, $La=i[h,a]$ only, with $h=\adj{h}\in M_{d}$.

Taking the Laplace transform of \eqref{nonmarkovian} one finds:
\begin{equation}\label{3.4}
(id\;p-L-\hat{L}_{p})\hat{A}_{p}=id.
\end{equation}
The equality before implies that both relations below:
\begin{equation}\label{3.5}
\hat{A}_{p}=(p\;id-L-\hat{L}_{p})^{-1}
\end{equation}
and
\begin{equation}\label{3.6}
\hat{A}_{p}(id-L-\hat{L}_{p})^{-1}=id,
\end{equation}
hold.

It follows from \eqref{3.6}  that equation \eqref{nonmarkovian} can also be written in the form
\begin{equation}\label{3.7}
\frac{dA_{t}}{dt}=A_{t}L+\int_{0}^tdsA_{s}L_{t-s},
\end{equation}
and consequently, the  dual dynamics becomes:
\begin{equation}
\frac{d\conj{A}_{t}}{dt}=-L\conj{A}_{t}+\int_{0}^tds\conj{L_{t-s}}\conj{A}_{s}.
\end{equation}
This means that in the case of non-Markovian master equations there is an analogy to the Markovian case.

To find conditions on $L$ and $L_{t}$ that ensure that the time evolution $A_{t}$ resulting from \eqref{nonmarkovian} is completely positive for all $t\geq 0$ is the fundamental problem of non-Markovian master equations. The main result of the current paper can be summarized in the following theorem.
\begin{thm}
Let us suppose that $A_{t}$ is the solution of the equation \eqref{nonmarkovian}, where $L_{t}$ has the form
\begin{equation}\label{3.9}
L_{t}=B_{t}+Z_{t},
\end{equation}
where $B_{t}\in\cpos{M_{d}}$ for all $t\geq 0$,
\begin{equation}\label{3.10}
Z_{t}a=-\frac{1}{2}\set{B_{t}(\1_{d}),a}+i[h_{t},a],
\end{equation}
and $h_{t}=\adj{h}_{t}$, then $A_{t}$ is completely positive for all $t\geq 0$ if the solution of the normalization equation
\begin{equation}\label{3.11}
\frac{dN_{t}}{dt}=LN_{t}+\int_{0}^tdsZ_{t-s}N_{s},
\end{equation}
with the initial condition $N_{0}=id$, is completely positive for all $t\geq 0$.
\end{thm}
\begin{pf}
It follows from \eqref{nonmarkovian} that the Laplace transform $\hat{A}_{p}$ of $A_{t}$ is given by the formula
\begin{equation}\label{3.12}
\hat{A}_{p}=(id\;p-L-\hat{Z}_{p}-\hat{B}_{p})^{-1},
\end{equation}
and satisfies the equation

\begin{equation}\label{3.13}
\hat{A}_{p}=(id\;p-L-\hat{Z}_{p})^{-1}+(id\;p-L-\hat{Z}_{p})^{-1}\hat{B}_{p}\hat{A}_{p}.
\end{equation}
It follows from \eqref{3.13} and \eqref{3.11} that \eqref{nonmarkovian} can be written in the form:
\begin{equation}\label{3.14}
A_{t}=N_{t}+\int_{0}^tdu\int_{0}^{t-u}dsN_{t-u-s}B_{u}A_{s}.
\end{equation}
If $N_{t}$ is completely positive for all $t\geq 0$, then iterating \eqref{3.14} it is easy to see that $A_{t}$ is completely positive as well for all $t\geq 0$, since $B_{t}\in\cpos{M_{d}}$, provided the iteration procedure converges.
\end{pf}

In order to analyze the problems related to the solution of the normalization equation let us consider the non-Markovian master equation of the form
\begin{equation}\label{3.15}
\frac{dA_{t}}{dt}=\int_{0}^tds k(t-s)(B_{t-s}-id)A_{s},
\end{equation}
where $B_{t}\in\cpos{M_{d}}$ and $B_{t}(\1_{d})=\1_{d}$ for all $t\geq 0$, and $k(t)\geq 0$.

The normalization equation takes the form
\begin{equation}\label{3.16}
\frac{dN_{t}}{dt}=-\int_{0}^tds k(t-s)N_{s},
\end{equation}
with the initial condition $N_{0}=id$.

The solution of \eqref{3.16} has the form
\begin{equation}\label{3.17}
N_{t}=f(t)id,
\end{equation}
where $f(t)$ satisfies the equation

\begin{equation}\label{3.18}
\frac{df(t)}{dt}=-\int_{0}^tds k(t-s)f(s),
\end{equation}
and $f(0)=1$.

As a particular case, let us choose $k(t)$ in the Lidar-Shabani form, cf. \cite{Shabani:2005gb}, ie.,
\begin{equation}\label{3.19}
k(t)=\kappa^2 e^{-2\kappa\gamma t}.
\end{equation}
In this case one easily finds
\begin{equation}
f(t)=\begin{cases}\label{3.20}
      & e^{-\kappa\gamma t}\left[\cos (\kappa t\sqrt{1-\gamma^2})+\frac{\gamma}{\sqrt{1-\gamma^2}}\sin (\kappa t\sqrt{1-\gamma^2})\right],\;\text{if $0\leq\gamma<1$ }, \\
      & e^{-\kappa t}(1+\kappa t),\;\text{if $\gamma=1$},\\
      &e^{-\kappa\gamma t}\left[\cosh \kappa t\sqrt{\gamma^2-1})+\frac{\gamma}{\sqrt{\gamma^2-1}}\sinh (\kappa t\sqrt{\gamma^2-1})\right],\text{if $\gamma>1$}.
\end{cases}
\end{equation}
It follows from \eqref{3.17} that $N_{t}$ is completely positive if and only if $f(t)\geq 0$ for all $t\geq 0$ for all $t\geq 0$, and \eqref{3.20} shows that $f(t)\geq 0$ for all $t\geq 0$ if and only if $\gamma\geq 1$.

The above example clearly indicates that the structure of non-Markovian master equations is much more complicated than the Markovian ones.

\section{Modified non-Markovian master equations}
The time evolution (in the Heisenberg picture) is given by the family of maps $A_{t}:M_{d}\to M_{d}$, $t\geq 0$, such that $A_{t}\in\vN{B}^+(M_{d})$, for all $t\geq 0$, (complete positivity condition), $A_{t}(\1_{d})=\1_{d}$ for all $t\geq 0$, (normalization condition) and $A_{0}:=\lim_{t\downarrow 0}A_{t}=id$. In section 3 it has been shown that if $A_{t}$ satisfies equation \eqref{nonmarkovian}, then the normalization condition can be imposed with no trouble. Indeed, if \eqref{3.2}, \eqref{3.9} and \eqref{3.10} are satisfied, then the normalization condition is trivially fulfilled. On the other hand, complete positivity of $A_{t}$ leads to complete positivity of solutions to the normalization equation \eqref{3.10} which is a very difficult problem. However one can circumvent the above difficulty in the following manner. Let $V_{t}$, $t\geq 0$ be the family of complete positive maps on $M_{d}$ such that $\lim_{t\to 0}V_{t}=id$. If $V_{t}(\1_{d})>0$ for all $t\geq 0$, then the maps $A_{t}$, $t\geq 0$, defined as
\begin{equation}\label{4.1}
A_{t}(a)=V_{t}(\1_{d})^{-1/2}V_{t}(a)V_{t}(\1_{d})^{-1/2},
\end{equation}
are completely positive and normalized.

Let $V_{t}$, $t\geq 0$ be the solution of the following modified non-Markovian master equation:
\begin{equation}\label{4.2}
\frac{dV_{t}}{dt}=PV_{t}+\int_{0}^tds B_{t-s}V_{s},
\end{equation}
 with the initial condition $\lim_{t\to 0}V_{t}=id$, where $P$ is a completely positive map and $B_{t}\in\vN{B}^+(M_{d})$ for all $t\geq 0$.

 The resolvent of \eqref{4.2},
 \begin{equation}\label{4.3}
\widehat{V}_{p}=(id\;p-P-\widehat{B}_{p})^{-1},
\end{equation}
satisfies the equation
\begin{equation}\label{4.4}
\widehat{V}_{p}=\left(id\;p-P\right)^{-1}+\left(id\;p-P\right)^{-1}\widehat{B}_{p}\widehat{V}_{p},
\end{equation}
which is the integral form of \eqref{4.2}.

Iteration of \eqref{4.4} yields that $V_{t}$ is completely positive since $\exp (tP)$ and $B_{t}$ are completely positive.

If the solution of \eqref{4.2} satisfies the condition $V_{t}(\1_{d})>0$ for all $t\geq 0$, then $A_{t}$, $t\geq 0$, defined through \eqref{4.1} gives the correct time evolution, i.e., it is completely positive and normalized.

The above approach contains as a special case the semigroup form of the dynamics. Let us consider the equation
\begin{equation}\label{4.5}
\frac{dV_{t}}{dt}=LV_{t}+\lambda^2\int_{0}^tds\;e^{(t-s)L}V_{s},
\end{equation}
where $L$ is the generator of a completely positive semigroup. One easily finds the solution of \eqref{4.5} which is of the form:
\begin{equation}
V_{t}=\cosh (\lambda t)e^{tL},
\end{equation}
and $V_{t}({\bf 1}_d)={\bf 1}_d\cosh (\lambda t)$. The corresponding
normalized evolution $A_{t}$ has the form
\begin{equation}
A_{t}=e^{tL},
\end{equation}
that is, it is a semigroup.

\def\cprime{$'$} \def\cprime{$'$}

\end{document}